\documentclass[a4paper,twocolumn,11pt,accepted=2025-01-15]{quantumarticle}
\pdfoutput=1
\usepackage[utf8]{inputenc}
\usepackage[english]{babel}
\usepackage[T1]{fontenc}
\usepackage{amsmath}
\usepackage{hyperref}
\usepackage{dsfont}  
\usepackage{subcaption} 
\usepackage{upgreek}  
\usepackage{bm}

\usepackage[numbers,sort&compress]{natbib}

\usepackage{xcolor}
\definecolor{quantumviolet}{HTML}{53257F} 
\hypersetup{
    colorlinks = true,
    linkcolor = quantumviolet,
    citecolor = quantumviolet,
    filecolor = quantumviolet,      
    urlcolor = quantumviolet,
}

\begin{document}

\title{Imperfect quantum networks with tailored resource states}

\author{Maria Flors Mor-Ruiz}
\email{Maria.Mor-Ruiz@uibk.ac.at}
\affiliation{Institut f\"ur Theoretische Physik, Universit\"at Innsbruck, Technikerstra{\ss}e 21a, 6020 Innsbruck, Austria}
\orcid{0000-0003-4921-5929}
\author{Julius Walln{\"o}fer}
\orcid{0000-0002-4837-2757}
\affiliation{Dahlem Center for Complex Quantum Systems, Freie Universit\"at Berlin, Arnimallee 14, 14195 Berlin, Germany}
\author{Wolfgang D\"ur}
\affiliation{Institut f\"ur Theoretische Physik, Universit\"at Innsbruck, Technikerstra{\ss}e 21a, 6020 Innsbruck, Austria}
\orcid{0000-0002-0234-7425}
\maketitle
\begin{abstract}
Entanglement-based quantum networks exhibit a unique flexibility in the choice of entangled resource states that are then locally manipulated by the nodes to fulfill any request in the network. Furthermore, this manipulation is not uniquely defined and thus can be optimized. We tailor the adaptation of the resource state or pre-established entanglement to achieve bipartite communication in an imperfect setting that includes time-dependent memory errors. In this same setting, we study how the flexibility of this approach can be used for the distribution of entanglement in a fully asymmetric network scenario. The considered entanglement topology is a custom one based on the minimization of the required measurements to retrieve a Bell pair. The optimization of the manipulation and the study of such a custom entanglement topology are performed using the noisy stabilizer formalism, a recently introduced method to fully track noise on graph states. We find that exploiting the flexibility of the entanglement topology, given a certain set of bipartite requests, is highly favorable in terms of the fidelity of the final state.
\end{abstract}

\maketitle

\section{Introduction} \label{sec:Introduction}
Quantum communication is at the forefront of upcoming quantum technologies, representing a cutting-edge approach to secure and efficient data transmission. The overall goal in this field is to build a quantum internet \cite{Kimble2008, wehner_internet, Azuma2021, Azuma2023, CacciapuotiInternet2020, Illiano2022, CacciapuotiWhen2020} that would allow the distribution of quantum entanglement between any two or more points on the planet. Said entanglement is a crucial resource to enable applications in different areas such as quantum cryptography \cite{Gisin2002, ShorSimple}, quantum sensing \cite{Eldredge2018, Sekatski2020}, and distributed quantum computing \cite{CiracDistributed, Cacciapuoti2020}. In particular, the benefits of the use of multipartite entanglement is a topic of notable interest in terms of the applications that use multipartite entangled states, e.g., conference key agreement \cite{Murta2020Quantum, hahn_anonymous}, (multipartite) quantum secret sharing \cite{markham_graph, Hillery1999}. However, it is also relevant to assess the benefits of the use of multipartite entanglement at any part of the distribution process. For example, entanglement-based quantum networks \cite{pirker_modular_2018, pirker_quantum_2019, miguel-ramiro_optimized_2021, morruiz2023influence} and measurement-based quantum computation \cite{RaussendorfOneWay, RaussendorfMeasurement, briegel_measurement-based_2009} are instances that use multipartite entangled states as resource states. 

Recently, in \cite{mor_noisy} an efficient method to describe noisy graph transformations has been found, the so-called noisy stabilizer formalism. In the framework of quantum networks, this is extremely useful since many protocols are formulated with graph states \cite{hein_multiparty_2004, hein_entanglement_2006} or employ resource states which can be described as a graph state \cite{hahn_quantum_2019, HahnLimitations2022, Meignant2019, Mannalath2023, Fischer2021}. Furthermore, the study of the impact of noise in the different stages of these applications is relevant to simulate and study the challenges faced in experimental realizations.

In this work, we focus on entanglement-based quantum networks. The central idea for this approach is the distribution of a multipartite entangled state across spatially separated network nodes, which is then locally manipulated to achieve tasks for the network, e.g., establishing bipartite quantum communication between two nodes. This approach is motivated by the improvement of quantum memories in recent years \cite{zhong2015optically, young2020half, wang2021single, Picken_2019} and corresponds to a top-down approach to quantum networks, as the entanglement is established before the request is known. This pre-establishment of the resource state allows for (i) short waiting times for the network users, since the resource state only needs to be locally manipulated to fulfill a network request and (ii) flexibility not present in bottom-up approaches to quantum networks \cite{matsuzaki_2010, vanmeter_recursive, epping2016large, pirandola2019end, wallnofer_simulating_2022, kozlowski_2019, coutinho2022robustness, Bugalho2023distributing}. In particular, here we use the fact that the distribution of this resource state or entanglement topology does not need to match the structure of the network to tailor the entanglement topology to a certain set of possible requests in the network. 

Notably, we consider an imperfect setting that is comprised of noise on the distribution of resource states, memory errors due to the storage of entanglement, and lastly the manipulation via noisy operations. Then, two protocols to achieve bipartite communication in the network that consider different node capacities are established. Next, we analyze the behavior of the network under this imperfect setting and the chosen protocol. This analysis, mostly focused on the role of memory errors, entails: 
\begin{itemize}
    \item The optimization of the manipulation of the resource state.
    \item The study of the impact of asymmetries in both memory quality and positioning of the nodes.
    \item The investigation of regimes where the network can benefit from the customization of the entanglement topology based on a set of possible network requests.
\end{itemize}

The paper is organized as follows. In Sec.~\ref{sec:Concepts} we provide the necessary concepts on entanglement-based quantum networks, graph states, and their manipulations, as well as the noisy stabilizer formalism and the approach we consider for the time-dependencies in the manipulation, the so-called \textit{one-shot approach}. In Sec.~\ref{sec:Setting} we discuss the setting, which includes the used noise model and the resource and target states we consider. Next, in Sec.~\ref{sec:Protocols}, we establish two network protocols to achieve the proposed target state considering the corresponding timings and communication of classical information. Then, we analyze the protocols under the considered setting in Sec.~\ref{sec:Results}, which includes the optimization of the manipulation of the resource state for a symmetric scenario, the introduction of asymmetries and their impact w.r.t. noise, followed by the analysis of a tailored resource state in an asymmetric scenario. Lastly, we summarize and conclude in Sec.~\ref{sec:Conclusion}.

\section{Concepts} \label{sec:Concepts}
\subsection{Entanglement-based quantum networks}
Entanglement-Based Quantum Networks (EBQNs) \cite{pirker_modular_2018, pirker_quantum_2019, miguel-ramiro_optimized_2021, morruiz2023influence} correspond to a top-down approach to quantum networks. They utilize multipartite entangled states which are locally manipulated to fulfill network tasks on demand. Three phases can be distinguished in these networks \cite{pirker_quantum_2019}, 
\begin{enumerate}
    \item Dynamic phase: A set of entangled states, both multipartite and bipartite, are generated and distributed among the network nodes when the network is idle. We refer to the distributed entanglement as the resource state.
    \item Static phase: After establishing the resource state, it is stored until there is a network request (e.g., a bipartite connection between two network nodes). In order to minimize memory resources, states with minimal required storage are preferred \cite{miguel-ramiro_optimized_2021}.
    \item Adaptive phase: Once the users of the network have a request, the nodes manipulate locally (assisted by classical communication) the resource state to achieve this request or target state, without further using quantum communication in the network.
\end{enumerate}
Long waiting times for the network users are avoided due to the pre-generated entanglement, which in turn requires high-quality quantum memories in the network devices for long-term storage. Note that this entanglement allows for flexibility in the network requests, since a resource state can be optimized to be able to fulfill many different requests.

\begin{figure}[h]
    \centering
    \includegraphics[width=\columnwidth]{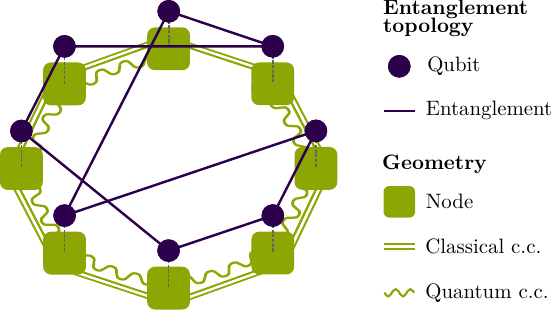}
    \caption{Schematic depiction of the different structures of an EBQN, where c.c. is short for communication channels. Note that the classical c.c., the quantum c.c. and the entanglement do not exactly match.}
    \label{fig:ebqn}
\end{figure}

One central feature of EBQNs is that the structure of the communication channels does not necessarily need to match the distribution of the entanglement of the resource states. For our purposes, we distinguish the \textit{geometry} and the \textit{entanglement topology} of a network. The former denotes the positioning of the nodes of the network and the communication channels, both classical and quantum. The quantum ones are mostly used to generate and distribute the resource state in the dynamic phase, and thus, are most probably located between neighboring nodes. However, the classical ones are used in the adaptive phase and might not coincide with the quantum ones. Conversely, the entanglement topology refers to the entanglement structure of the qubits of the resource state. There are many possible choices for a resource state that can fulfill a given set of requests. Therefore, one can customize the resource state to optimize the performance of the network. Fig.~\ref{fig:ebqn} graphically represents the different structures and channels in an EBQN. Note that the processes in the dynamic phase are highly dependent on hardware, and thus, can differ greatly, e.g., see \cite{Meignant2019, Fischer2021, PhysRevA.107.012609, Bugalho2023distributing, PhysRevA.100.032310, Wang_2009, PRXQuantum.4.040323} for graph, GHZ, and W states distribution.

\subsection{Graph states}
Graph states \cite{hein_multiparty_2004, hein_entanglement_2006} have been identified as suitable candidates for the resource states, as they allow for the transformation to other entangled states via local operations and classical communication. These states form a subclass of genuinely multipartite entangled states where a state can be associated with a graph $G=(V, E)$. Qubits are represented by the set of vertices $V$, and the set of edges between vertices $E$ denotes entangling operations between two qubits, in particular, controlled-$Z$ gates. The corresponding graph state is
\begin{equation}
    \left|G\right\rangle = \prod_{(v,w)\in E}CZ^{(v,w)}\left|+\right\rangle^{\otimes V},
\end{equation}
where $\left|+\right\rangle$ is the +1 eigenstate of $\sigma_x$. This graph state is also defined as the unique $+1$ eigenstate of stabilizers 
\begin{equation}
    K_v = \sigma_x^{(v)}\prod _{w\in \mathcal{N}_v}\sigma_z^{(w)}
\end{equation}
for all $v\in V$, where $\mathcal{N}_v$ is the set of vertices adjacent to $v$, the neighborhood of $v$. 

Graph states can be manipulated and transformed by local Clifford operations and Pauli measurements \cite{hein_entanglement_2006, hein_multiparty_2004}, where the former is a set of unitary quantum operations that map stabilizer states to stabilizer states. For example, the \textit{local complementation} is a local Clifford operation that transforms the entanglement structure of a graph state such that given some vertex $v$, the sub-graph of the neighborhood of $v$ is inverted. The local Pauli measurements lead to another graph state (up to local corrections that depend on the measurement basis and outcome) where the measured qubit is disconnected. Measurement in the $\sigma_z$ basis simply deletes all edges associated with the measured vertex, it is also referred to as \textit{vertex deletion}. The measurement in the $\sigma_y$ basis acts as a local complementation applied to the measured qubit followed by the deletion of it. The measurement in the $\sigma_x$ basis acts as a local complementation on a neighboring qubit of the measured one, then a local complementation on the measured qubit, followed by the deletion of it and lastly repeating the local complementation on the neighboring qubit of the first step. 

\subsection{Noisy stabilizer formalism}
To describe and analyze noise acting on graph states and their manipulation, the Noisy Stabilizer Formalism (NSF) was introduced in \cite{mor_noisy}. This method allows one to accurately track the full description of Pauli-diagonal noise and, for certain important noise models such as local initial Pauli noise (e.g., see Sec.~\ref{ssec:noise:model}), scales linearly in the number of qubits of the initial state, rather than exponentially. This is possible because it independently updates the graph states and noise operators, which in turn, reveals the contributions of different sources of noise on the final state. These updates are based on the commutation relations between Pauli noise operators and manipulation operators and are referred to as \textit{update rules}. Once the noise operators and the graph state are updated, their size must match. Thus, if the target state is small, one can efficiently apply the noise maps to it and retrieve the final mixed state, thereby avoiding computation with large density matrices. Besides the size of the target state, the number of terms in each of the individual noise maps and the vertex degree of the considered graph also affects the efficiency of the method. The vertex degree is relevant due to the fact that noise operators acting in a single qubit might spread to a noise operator acting on several qubits, potentially leading to more required noise map updates going forward.

This is in contrast to approaches that use the efficient classical simulation of Clifford Circuits (e.g., \cite{Aaronson2004, Anders2006, Bravyi2019simulationofquantum, cirq_developers_2023_10247207, gadi_aleksandrowicz_2019_2562111, Gidney2021stimfaststabilizer}) and handle noise processes in a Monte Carlo fashion by randomly inserting additional gates. This type of approach allows to sample from the output distribution; however, it does not provide a full description of the final state. Compared to other methods of tracking noise transformation under Clifford operations \cite{janardan2016analytical, Miller2018}, the NSF method, in particular, allows tracking individual noise contributions instead of a combined error probability vector and uses a formulation that is inherently based on graph states. In this work, we use the NSF with local Pauli noise as a noise model, which while highly structured, does not correspond to a sparse error probability vector. An implementation of this formalism for graph states has been developed using Python and it is publicly available on \cite{nsf_github}.

\subsection{One-shot approach}
So far, we have described the manipulations of graph states in terms of the associated graph transformations. However, in this formulation, correction operations need to be applied after each measurement to recover a true graph state after each step. This introduces a strict time order in which the measurements and corrections need to be performed, as the precise correction operations depend on the outcome of the measurements. This is irrelevant in the analysis of noiseless settings with no communication delay between qubits but crucial when considering noise and/or time dependencies. 

Since the graph state transformations only consist of Clifford operations and Pauli measurements, it is possible to formulate an equivalent approach that performs a set of Pauli measurements and then applies a single round of correction operations at the end. That is, for every sequence of graph transformations, a so-called \textit{strategy} in the NSF, there exists a \textit{measurement pattern} of Pauli measurements for which the order of measurements is irrelevant. We refer to this as the one-shot approach. 

This approach is analogous to the perspective found in measurement-based quantum computation \cite{RaussendorfOneWay, RaussendorfMeasurement, briegel_measurement-based_2009}, which also considers measurements on a graph state that acts as a resource. Despite the measurement outcomes being random, the final state or outcome of the computation can be deterministically retrieved independently of the order and outcomes of the measurements, as long as only Pauli measurements are used. 

Two key insights allow this. The correction operations, which are necessary to retrieve a graph state after a measurement, are always local Clifford operators, i.e., local basis changes of the affected qubits. So, instead of performing the correction operation, one can simply change the basis of the measurement accordingly, as due to the properties of the Clifford group, any local Pauli measurement is always mapped to the same or a different local Pauli measurement. Furthermore, the local Clifford operators associated with different measurement outcomes are the same up to local Pauli operators (see, e.g., \cite{hein_entanglement_2006}), which can potentially swap the sign of a Pauli measurement, i.e., they can change the interpretation of the measurement outcome, but not the basis in which one measures. In Appendix~\ref{a:MeasurementPatterns}, we give an explicit computation of these measurement patterns and their equivalence with the strategies, we also analyze how this approach can be simulated using the NSF.

\section{Setting}\label{sec:Setting}
The work presented here is in the framework of EBQNs. In particular, the focus is on the adaptive phase, where the resource state has already been distributed and stored for some time, leading to a noisy resource state. This phase is triggered by a request in the network, such that the noisy resource state is manipulated locally by the nodes via imperfect operations to fulfill it.

We consider a network consisting of $N$ nodes where a cluster state is distributed as a minimal-storage resource state. In particular, we study a single bipartite communication as the request, as it is a useful target state in quantum communication tasks, such as quantum key distribution \cite{ekert_1991}. This target state, the bipartite graph, is a Bell state up to a local basis change. 

\subsection{Noise model}\label{ssec:noise:model}
We consider two kinds of local noise, depolarizing and dephasing. The noise arising from the dynamic and adaptive phases, which includes the preparation, distribution, and manipulation using imperfect local operations of the resource state, is described by depolarizing noise that acts on each qubit of the resource state independently. Additionally, the storage of the qubits in the static phase is modeled via local time-dependent dephasing noise. The depolarizing noise acting on qubit $v$ is defined as
\begin{equation}
    \mathcal{D}_v\rho = \left(1 - \frac{3p}{4}\right)\rho +\frac{p}{4}\sum_{i\in\{x,y,z\}}\sigma_i^{(v)}\rho \,\sigma_i^{(v)},
\end{equation}
where $p$ is the probability that qubit $v$ is depolarized. The time-dependent dephasing noise acting on qubit $v$ is defined as 
\begin{equation}\label{eq:dephasing:noise}
    \mathcal{Z}_v\rho = \left[1-q_v(t)\right]\rho + q_v(t)\sigma_z^{(v)}\rho\,\sigma_z^{(v)},
\end{equation}
where $q_v(t)$ denotes the time-dependent noise probability of the storage. Following \cite{hein_entanglement_2005}, we define this coefficient as $q_v(t)=\left(1 - e^{-t/T_v}\right) / 2$, where $t$ is a certain time in the network, and $T_v$ is the dephasing time, which we use to describe the quality of the memories in the network. The qubits in the network are expected to be affected by this noise for different amounts of time, e.g., until being measured or while waiting for all necessary information to arrive, which naturally depends on the protocol that is used.  Nevertheless,  Eq.~\eqref{eq:dephasing:noise} is valid for any time interval $t$. Therefore, each qubit is subject to $\mathcal{E}_v= \mathcal{D}_v\mathcal{Z}_v$, then an $N$-qubit noisy resource state is 
\begin{equation}
\mathcal{E}_1 \mathcal{E}_2 \cdots \mathcal{E}_N \left|G\rangle\langle G\right|, 
\end{equation}
which is then manipulated into a noisy version of the desired target state.  

The use of single-qubit noise models in this scenario is reasonable as we assume that during the static phase, the qubits are stored in spatially separated nodes. Moreover, one candidate method for resource generation and distribution in EBQNs is the use of entanglement purification, which leads to a nearly local noise model \cite{wallnoefer_meas}. Furthermore, local noisy measurements are accurately modeled using local depolarizing noise followed by a perfect measurement. Overall, the depolarizing noise model is considered general enough to accurately describe uncorrelated noise effects and can be understood as a worst-case model for single-qubit noise \cite{dur_standard}. The use of time-dependent dephasing noise is widely used to describe the decoherence process of a quantum memory \cite{Razavi2009, luong2016overcoming, wallnofer_simulating_2022}, and there are indications that dephasing errors are the limiting factor in diamond-based nuclear-spin memories \cite{Kalb2018, Pompili2021}. Importantly, the two noise models we consider are Pauli-diagonal noise channels, such that their impact on graph states can be efficiently described using the NSF.

\subsection{Entanglement topology and manipulation}
As mentioned in Sec.~\ref{sec:Concepts}, the choice of the resource state in EBQNs must minimize storage; therefore, we choose a state with a single qubit per node. In \cite{morruiz2023influence}, several structures under a noise model consisting only of single-qubit depolarizing noise that ignores memory decoherence and timing and spacing effects were studied as resource states to achieve a Bell pair with the conclusion that high-dimensional cluster states are good candidates for this task. So, in this work consider an $N$-qubit $d$-dimensional cluster state as a resource state with a noise model that incorporates the same depolarizing noise and additional memory decoherence. 

\begin{figure}[h]
    \centering
    \includegraphics[width=\columnwidth]{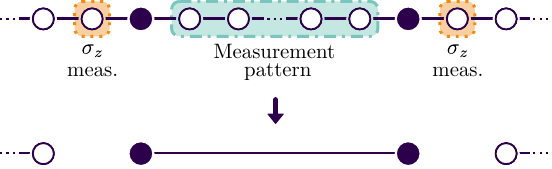}
    \caption{Manipulation of a 1D cluster into a Bell pair. The circles are qubits, the solid lines represent the entanglement between qubits and the dotted lines represent the extension of the cluster following the presented structure, so only part of a large cluster is explicitly drawn. At the top, the initial situation is depicted, where the target qubits (filled circles) have been chosen. Also, the corresponding manipulation is established, which consists of the measurement in the $\sigma_z$ basis of the outer neighbors and applying a measurement pattern to the inner neighbors. At the bottom, the manipulation has been performed and the resulting Bell pair and the remaining graph state are presented.}
    \label{fig:manipulation}
\end{figure}

To achieve a Bell pair from a cluster state, first, we choose the path to connect the two target qubits. We aim for the shortest one as the results in \cite{morruiz2023influence} show that the number of measurements required in the manipulation directly impacts the noise of the Bell pair. There are two kinds of qubits involved in the manipulation, the ones surrounding the path and the ones inside the path. We refer to them as \textit{outer} and \textit{inner} neighbors, respectively. The outer neighbors are measured in the $\sigma_z$ basis, as one wants to disconnect the target qubits and the inner neighbors from the rest of the cluster. This leads to a one-dimensional cluster with the targets at the ends and the inner neighbors in between. Then, for the measurement of the inner neighbors, multiple sets of measurement patterns achieve the target Bell pair. The choice of such a measurement pattern leads to different noise maps acting on the target Bell pair, which in turn may result in different fidelities. Therefore, we do not consider one fixed measurement pattern, but we rather analyze the impact of choosing different measurement patterns to pick an optimal one for the particular setup. In Fig.~\ref{fig:manipulation}, we graphically represent the manipulation for the one-dimensional case.

\subsection{One-dimensional cluster as a building block}
When manipulating a cluster state into a Bell pair, after isolating the path, one is left with a 1D cluster with the target qubits at the ends, where the inner neighbors are measured in the corresponding basis to obtain a Bell pair. The number of qubits surrounding the path in higher-dimensional clusters depends on the shape of the chosen path, but as in \cite{morruiz2023influence}, one can consider a straight-line path as a lower-bound approximation, as it has the maximum number of outer neighbors. This manipulation can be improved in terms of minimizing the number of required measurements as shown in \cite{hahn_quantum_2019} for the two-dimensional case, however, it can be used as a lower-bound scenario. Therefore, the 1D cluster is a well-suited building block for higher-dimensional clusters as resource states. Moreover, the use of path-finding algorithms is not necessary for the one-dimensional case. In the next results, we consider a 1D cluster as a resource state.

Importantly, in \cite{morruiz2023influence} it is shown that cluster states of higher dimension under local depolarizing noise have better behavior as the size of the network increases. To analyze this noise on such clusters, in \cite{morruiz2023influence}, the straight-line approximation is taken, such that the impact of the depolarizing noise on the target Bell pair increases linearly with dimension, keeping the same structure as for the one-dimensional case, but exponentially with the number of inner neighbors. Nevertheless, in this work, we are also considering time-dependent dephasing noise. The measurement of the outer neighbors is insensitive to the dephasing noise, as they are always measured in the $\sigma_z$ basis. Thus, to understand the impact of the time-dependent dephasing noise, on a high-dimensional cluster is sufficient to study the impact on the inner neighbors and target qubits. 

\begin{figure}[h]
    \centering
    \subfloat[Basic]{\includegraphics[width=0.5\columnwidth]{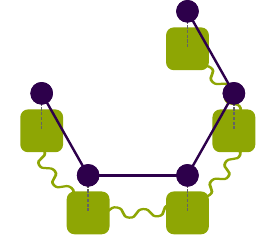}\label{fig:entanglement:topology:basic}} 
    \hfill
    \subfloat[Custom]{\includegraphics[width=0.5\columnwidth]{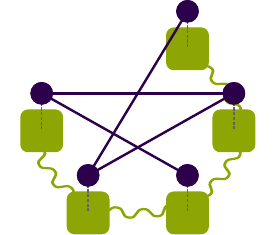}\label{fig:entanglement:topology:custom}} 
    \caption{Entanglement topologies of a 5-qubit 1D cluster. The squares are the nodes, the snake-like lines are the quantum communication channels, the circles are qubits and the solid straight lines represent the entanglement between qubits. The network nodes and the quantum communication channels in (a) and (b) are the same.}
    \label{fig:entanglement:topology}
\end{figure}

For a specific geometry, the distribution of the 1D cluster is not uniquely defined, such that it can be customized. For example, considering the process of generation of the resource state, we assume that the quantum communication channels are established between neighboring stations, so the most basic choice is to place the neighboring qubits in the 1D cluster in neighboring network nodes, then, entanglement does not have to be distributed over long distances. We refer to this distribution of the 1D cluster as the \textit{basic} entanglement topology, depicted in Fig.~\ref{fig:entanglement:topology:basic}. 

However, one can customize the entanglement topology with the targets of the network in mind. If the pairs of nodes that are most probable to request a Bell state are known, one can distribute the resource state such that these pairs have a direct entanglement link or a minimal number of qubits in between in the entanglement topology. Then, the number of measurements in the manipulation to achieve these pairs is minimized. An instance of this entanglement topology is depicted in  Fig.~\ref{fig:entanglement:topology:custom}. Later in this work, we analyze the advantages of having a target-oriented custom entanglement topology in terms of the fidelities of the target Bell pairs. Despite the possible advantages of minimizing the noise impact, such customization would require more advanced techniques for the entanglement generation process.

\section{Protocols}\label{sec:Protocols}
Here we establish protocols in the setting described above, taking into account the timings and the corresponding classical information that needs to be sent to achieve a request. In general, for the manipulation of the resource state one needs to account for delay times from (i) communicating the tasks, (ii) processing times of the corresponding nodes, and (iii) communicating measurement outcomes to the target nodes. 

Since, in this work, we discuss distributed graph states with qubits stored in imperfect quantum memories at spatially separated parties, the use of the one-shot approach is crucial. With this approach, measurements are performed as soon as possible, as there is no need to wait for any measurement outcome. Therefore, this avoids potential additional delays in network protocols, particularly if the geometry and entanglement topology do not match. We describe two protocols based on different structures and capabilities of the network nodes. Importantly, note that the following protocols only consider a single network request per resource state.

\subsection{Local protocol}
\begin{figure}[h]
    \centering
    \subfloat[Local]{\includegraphics[width=0.5\columnwidth]{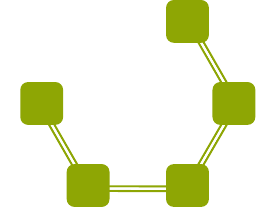}\label{fig:local}}
    \hfill
    \subfloat[Central]{\includegraphics[width=0.5\columnwidth]{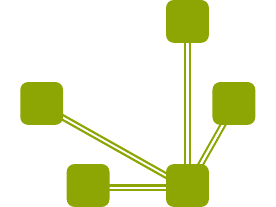}\label{fig:central}}
    \caption{Location of the classical communication channels for the local and the central protocols for a 5-node network. The squares are the nodes, which are the same for (a) and (b), and the double lines are the classical communication channels. The coordinator in (b) is the bottom-right node.}
    \label{fig:protocols}
\end{figure}

The local protocol we consider is characterized by decision-making nodes. We assume that all nodes have the same capacities and they organize the manipulation between themselves. The structure of the classical communication channels here follows a one-dimensional chain. So, each node has two classical communication channels that connect it to the two closest nodes, but the ends of the chain only have one, as depicted in Fig.~\ref{fig:local}. Such that the communication between nodes is done through this chain. Therefore, if the geometry and topology do not coincide, for a message to arrive at a certain node, it may have to pass through other nodes that might not be involved in the manipulation. 

For the nodes to implement this coordination, after the dynamic phase of the network, they agree on a measurement pattern to achieve a network request, a so-called \textit{rule set} using classical communication. Furthermore, all nodes know the geometry and the entanglement topology of the network, so each of them knows if it is involved in the manipulation, and what action it should take. Note that this rule set architecture is rather common in classical communication and has also been explored in the quantum regime, e.g., \cite{Matsuo2019}.

\begin{table}[h]
\centering
\resizebox{\columnwidth}{!}{
\begin{tabular}{p{\columnwidth}}
\hline \hline
\multicolumn{1}{c}{\textbf{Local Protocol}} \\ 
\hline
\textit{Input:} One of the target nodes, assume $a$, requires a Bell pair with $b$.
\begin{enumerate}
    \item Node $a$ forwards this request to the closest nodes and initializes the chain of measurements: 
    \begin{enumerate}
        \item Once the closest nodes receive the request, they decide if the message should be relayed to their other closest nodes or not. This decision depends on whether they are on the path of classical communication to one of the qubits in $\{q_i\}$, which are defined by the entanglement topology. 
        \item These nodes check the entanglement topology to see if their qubit belongs to $\{q_i\}$, and if so, consult the rule set to determine the basis of the measurement.
        \item This relay-measure procedure continues until no further relying is necessary. Meanwhile, the nodes that have performed a measurement send the outcome to the corresponding target nodes via the classical channels.
    \end{enumerate}
    \item The target nodes receive all outcomes. 
    \item $a$ and $b$ apply the corrections
\end{enumerate} 
\textit{Output: }$a$ and $b$ share a Bell pair. \\
\hline \hline 
\end{tabular}
}
\caption{Detailed steps of the local protocol. The target nodes to establish a Bell pair between are denoted by $a$ and $b$, and $\{q_i\}$ are the qubits that have to be measured to achieve the target.}
\label{tab:local:protocol}
\end{table}

In Table~\ref{tab:local:protocol}, we present a step-by-step description of the local protocol. Note that the time for all the necessary outcomes to arrive and then perform the correction might be different for each target qubit, such that one might have to wait for the other. In Appendix~\ref{a:delay:times:local}, we compute the delay times for all the involved qubits.

\subsection{Central protocol}
In contrast to the local protocol, consider another setup where one of the nodes has more capabilities than the rest of them. In this central protocol, we refer to this more powerful node as the coordinator, which is the only one that knows the geometry and entanglement topology of the network and is in charge of all decision-making in the network. Then, the rest of the nodes of the network can be simpler than they are for the local protocol.

A request enters the network through the coordinator who is in charge of organizing the manipulation to achieve the target. Therefore, the classical communication channels of this protocol are established from each node to the coordinator, following a star-like structure as depicted in Fig.~\ref{fig:central}. 

\begin{table}[h]
\centering
\resizebox{\columnwidth}{!}{
\begin{tabular}{p{\columnwidth}}
\hline \hline
\multicolumn{1}{c}{\textbf{Central Protocol}} \\ 
\hline
\textit{Input:} Arrival of the request at $C$.
\begin{enumerate}
    \item $C$ chooses the measurement pattern to achieve the target pair. 
    \item $C$ prepares the commands for $\{n_i\}$, which include
    \begin{enumerate}
        \item basis of the measurement ($\sigma_x$, $\sigma_y$, $\sigma_z$),
        \item order to perform the measurement,
        \item order to send back the outcome.
    \end{enumerate}
    \item $C$ sends the commands to $\{n_i\}$ classically, through the classical channel between $C$ and each party.
    \item Given the one-shot approach, as each of the qubits in $\{n_i\}$ receives the command, they perform it.
    \item When $C$ has received all the outcomes, the correction operators are computed and sent to $a$ and $b$. 
    \item $a$ and $b$ apply the corrections
\end{enumerate} 
\textit{Output: }$a$ and $b$ share a Bell pair. \\
\hline \hline 
\end{tabular}
}
\caption{Detailed steps of the central protocol. $C$ denotes the coordinator, $a$ and $b$ denote the target nodes to establish a Bell pair between, and $\{n_i\}$ are the nodes that have to perform a measurement to achieve the target.}
\label{tab:central:protocol}
\end{table}

In Table~\ref{tab:central:protocol}, we present a step-by-step description of the central protocol. Note that performing the commands in step 4, might not be a simultaneous action, as some nodes might receive the command later than others due to their longer distance to the coordinator. Also, step 6 is not necessarily simultaneous, so one of the target qubits might have to wait for the other one to apply the correction. In Appendix~\ref{a:delay:times:central}, we compute the delay times for all the involved qubits. 

\section{Simulation results}\label{sec:Results}
Here we present the results obtained from the simulation and optimization of specific scenarios of the setting presented in Sec.~\ref{sec:Setting} with the protocols described in Sec.~\ref{sec:Protocols}. For all results, we consider that classical information travels at the speed of light in optical fiber, $2\cdot 10^8$ m/s, and that all nodes have a processing time of 1 $\upmu$s to perform measurements on their qubit. The parameters that change in each studied scenario are the node positioning, the strength of the depolarizing noise, $p$, and the dephasing times, $T_v$. In Appendix~\ref{a:noise}, we detail how the noise of the measurement patterns is studied in the simulation.

\subsection{Optimal measurement pattern}\label{ssec:Optimal}
Our goal is to optimize the measurement pattern to manipulate a 1D cluster into a Bell pair. To do so, we consider a symmetric scenario, such that the distance between nodes is constant, all qubits are subject to depolarizing noise with the same $p$ and have the same dephasing time $T$. Moreover, the nodes are placed in a line, so they all have the same $x$ or $y$ coordinate. We assume a basic entanglement topology, meaning that neighboring qubits are placed in neighboring nodes.

We study manipulating an $N$-qubit 1D cluster into a Bell pair between the end nodes, so there are $N-2$ inner neighbors and no outer neighbors. All measurement patterns that retrieve a Bell pair are formed by $\sigma_y$ and $\sigma_x$, thus, there are $2^{N-2}$ of them\footnote{Measurements in the $\sigma_x \sigma_y$ plane also lead to a Bell pair. However, we do not consider them as including this would dramatically increase the number of possible measurement patterns, and then an exhaustive search over all of them would have an even higher computational cost. Also, the correction operator from these measurements is not Clifford, such that the state is no longer in the graph state basis and the use of the NSF for measurement patterns (see Appendix~\ref{a:noise}) is no longer valid.}. We exclude $\sigma_z$, as this would introduce a cut in the 1D cluster, preventing the final connection between the two target qubits.

We fix the inter-node distance to 15 km and we vary the depolarizing strength and the dephasing time as $p\in[0, 0.1]$ and $T\in[1, 100]$ ms. Given this parameter space, we perform an exhaustive search by computing the fidelity of the final Bell pair using all possible measurement patterns, and then we compare them. We only use the exhaustive search for $N\leq 12$, due to the costly computation.

We label the qubits and nodes as $\{1, \dots, N\}$, where the targets are 1 and $N$ and the inner neighbors are $\{2, \dots, N-1\}$. Importantly, in the local protocol, we assume that node 1 is the one starting the protocol, and in the central protocol, we assume that the coordinator is node $\left\lceil N/2\right\rceil$. We express a measurement pattern as $\boldsymbol{m}$, an $N-2$ length vector where $m_i\in\{\sigma_x, \sigma_y,\sigma_z\}$, and the $i^{\text{th}}$ vector element corresponds to the measurement basis on the $(i+1)^{\text{th}}$ qubit. 

The results obtained for small systems (up to $N\leq 12$) show a binary behavior of the optimal measurement pattern, such that we can distinguish two regimes. One where the depolarizing noise is dominant and another where the dephasing noise is dominant, which corresponds to low $T$. Each regime has an optimal measurement pattern that changes when entering the other regime. Importantly, there is an area in between where the two measurement patterns are optimal and equivalent, which we call the transitional regime. 

From the observed optimal patterns for small systems, we extrapolate the following measurement patterns for all $N$. If $N$ is even, in the depolarizing regime the optimal measurement pattern is $(\sigma_x, \sigma_x, \dots, \sigma_x)$, for both protocols. However, for the dephasing regime, the optimal measurement pattern is different. For the local protocol is $(\sigma_y, \sigma_y, \sigma_x, \dots, \sigma_x)$, and for the central protocol the optimal measurement pattern has the following elements:
\begin{equation}
    m_i =
    \begin{cases}
        \sigma_y \text{ if } i=\frac{N}{2}, \frac{N}{2}\pm 1,\\
        \sigma_x \text{ otherwise}.
    \end{cases}
\end{equation}
If $N$ is odd, for both the local and central protocol, the optimal measurement pattern is always $(\sigma_x, \sigma_x, \dots, \sigma_x)$. 

Notice that in the depolarizing regime, the optimal measurement pattern is the same for both protocols, whereas in the dephasing regime, there are some differences. This is because the two protocols differ in the delay times, so the final dephasing noise on the Bell pair differs in each protocol, but the final depolarizing noise is the same.

\begin{figure}[!t] 
    \centering
    \subfloat[Local protocol]{\includegraphics[width=\columnwidth]{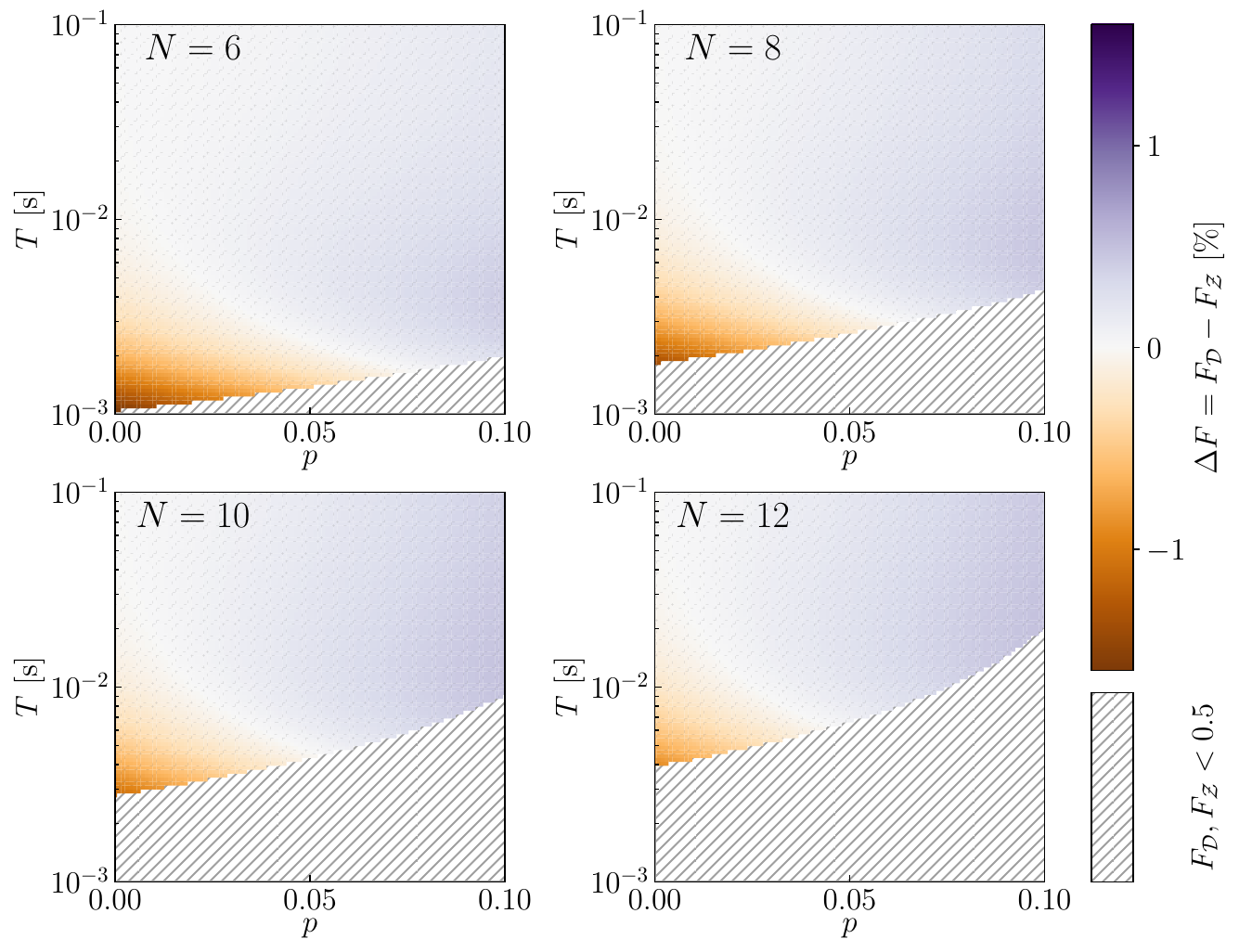}}
    \vfill
    \subfloat[Central protocol]{\includegraphics[width=\columnwidth]{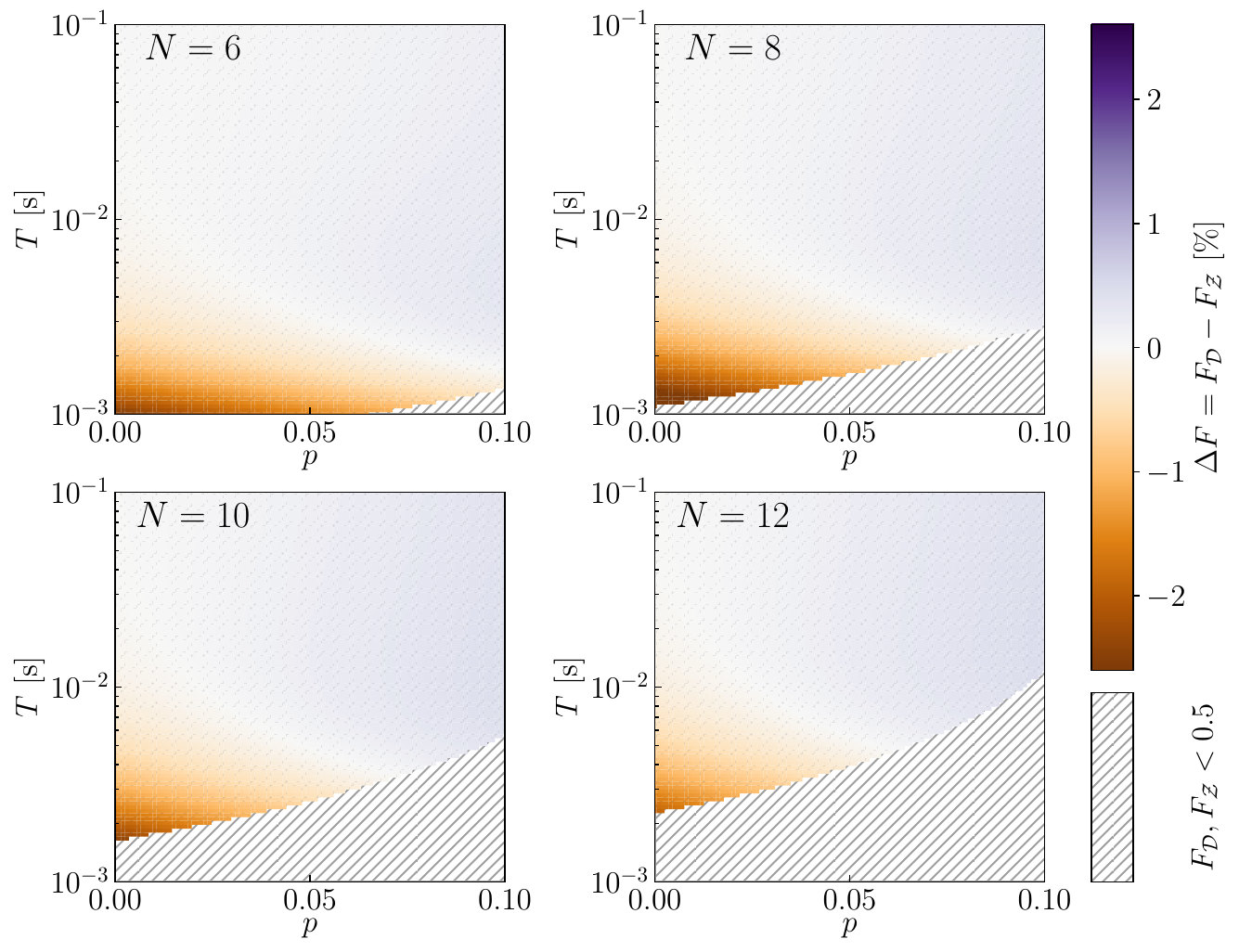}}
    \caption{Difference between the fidelities achieved using the optimal measurement pattern in the depolarizing regime ($F_{\mathcal{D}}$), and the dephasing regime ($F_{\mathcal{Z}}$), for both local and central protocols, with even $N$, in terms of the strength of the depolarizing noise $p$ and the dephasing time $T$. The striped area denotes that both fidelities $F_{\mathcal{D}}$ and $F_{\mathcal{Z}}$ are below 0.5, and the target Bell pair is no longer entangled.}
    \label{fig:optimal:pattern}
\end{figure}

In Fig.~\ref{fig:optimal:pattern}, we plot the difference between the fidelities achieved using the two optimal measurement patterns for the two protocols with even $N$. There we can clearly distinguish the two regimes. The depolarizing regime is mostly at the top-right part of each plot and the dephasing regime is for low values of $p$ and relatively low values of $T$. Between the two regimes, there is a white thin area that denotes the transitional regime, where there is no difference between fidelities. In Appendix~\ref{a:symmetric}, we present the achieved fidelities with the optimal measurement patterns. 

In a noiseless scenario, the choice between these measurement patterns does not cause any difference in the final Bell pair. However, here we observe that different measurement patterns lead to different fidelities. So, overall, the optimization of measurement patterns is necessary in terms of the noise impact, and even though we have seen that for this setting only a small selection of measurement patterns is relevant and their differences in terms of fidelity are at most of a $2\%$, this might change with other noise models.

\subsection{Introduction of asymmetry}
Here we introduce certain asymmetries to the previous symmetric scenario and we analyze their impact on the final pair. We once again consider that the $N$ nodes are placed in a line, and we assume a basic entanglement topology. The target state is a Bell pair between the end nodes of the $N$-qubit 1D cluster. Here we also label the qubits and nodes as $\{1, \dots, N\}$, such that in the local protocol, node 1 is the one starting the protocol, and in the central protocol, the coordinator is node $\left\lceil N/2\right\rceil$.

\subsubsection{Asymmetry in memory quality}
First, we introduce asymmetries in the dephasing times, which would represent having quantum memories of different qualities in the network. Thus, we consider that the dephasing time of one of the qubits ranges from $0.1$ ms to $100$ ms, and the remaining qubits have a fixed dephasing time of $100$ ms. All other parameters are symmetric, so we consider depolarizing noise with $p=1\%$ for all qubits, and inter-node distances of 15 km. Given this parameter space, we investigate the optimal measurement pattern for the specified parameters to find that the optimal is always $(\sigma_x, \dots, \sigma_x)$, which is in agreement with the results found for the symmetric scenario. 

Using this measurement pattern we study the effect of a faulty memory depending on where is it placed in the network. In general, a faulty memory in the network decreases the fidelity of the Bell pair, and if the faulty qubit is in a node with long delay times the impact on the fidelity is larger. For the local (central) protocol, the farther away a node is from node 1 (node $\left\lceil N/2\right\rceil$), the longer delay times it has. 

\begin{figure}[h] 
    \centering
    \subfloat[Local protocol]{\includegraphics[width=0.5\columnwidth]{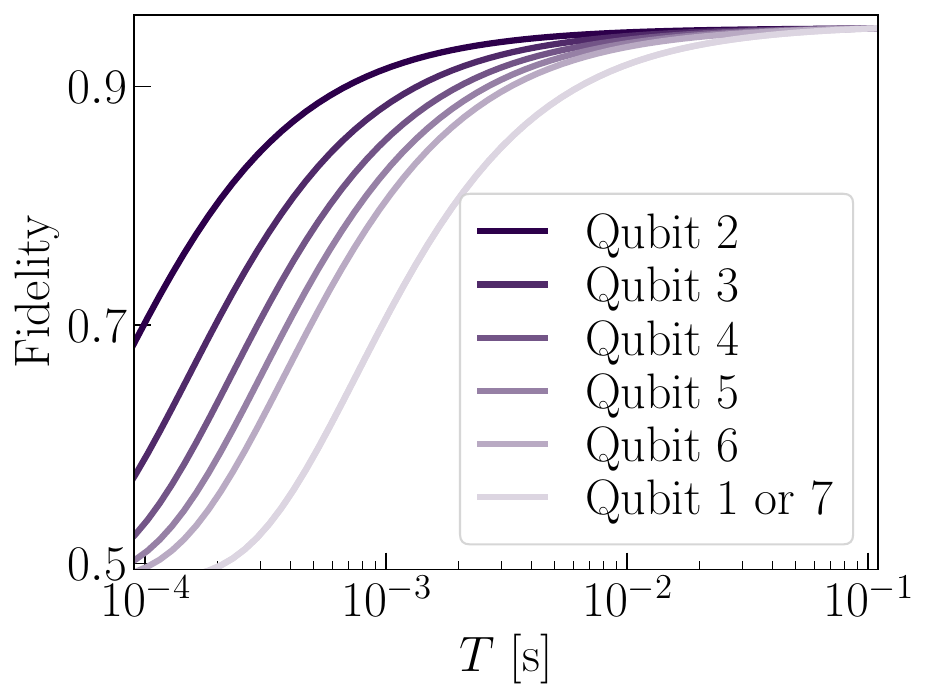}\label{fig:memory:asymmetry:local}}
    \subfloat[Central protocol]{\includegraphics[width=0.5\columnwidth]{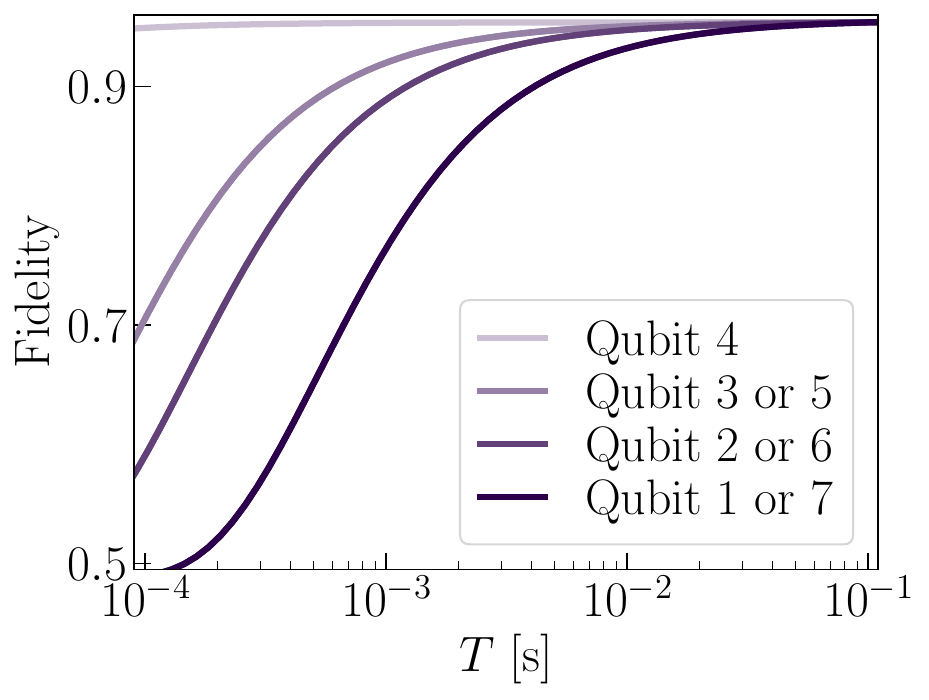}\label{fig:memory:asymmetry:central}}
    \caption{Fidelity of a Bell pair obtained from a 7-qubit 1D cluster in terms of the dephasing time of the faulty memory with the local or the central protocol. Each line denotes which is the faulty qubit. Lines that coincide are plotted in the same color.}
    \label{fig:memory:asymmetry}
\end{figure}

In Fig.~\ref{fig:memory:asymmetry}, we plot the fidelity of the final Bell pair in terms of the dephasing time of the faulty memory, for both protocols with $N=7$, where we observe the behavior described above. For the local protocol, a faulty memory in one of the target qubits (qubits 1 and 7 in Fig.~\ref{fig:memory:asymmetry:local}) has the same effect, as they have to wait for each other, furthermore, since their delay times are longer than the others the impact of their faulty memory is higher. For the central protocol, we see that if the coordinator (qubit 4 in Fig.~\ref{fig:memory:asymmetry:central}) has a faulty memory it does not affect the fidelity, as the qubit in the coordinator is only subject to the delay time related to the processing time. Also, qubits that have the same distance to the coordinator have equal delay times, such that the change in fidelity due to a faulty memory in these qubits is the same, as one can see in the overlap of lines in Fig.~\ref{fig:memory:asymmetry:central}.

\subsubsection{Asymmetry in position}
\begin{figure}[h]
    \centering
    \includegraphics[width=\columnwidth]{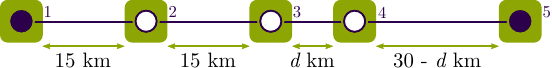}
    \caption{Shifting of the position of node 4 in a 5-node network. The squares are the nodes, the circles are qubits and the lines represent the entanglement between them. Filled circles denote target qubits and empty circles denote inner neighbors.}
    \label{fig:drawing:position:asymmetry}
\end{figure}

Now, we consider symmetry in the dephasing times but introduce asymmetries in the positioning of the nodes. In general, the inter-node spacing is 15 km, but the position of one of the qubits is shifted, such that the inter-node distance to the previous node is $d$ km and to the next node $30 - d$ km, and we consider $d\in[5, 25]$ km. This is depicted in Fig.~\ref{fig:drawing:position:asymmetry} for $N=5$, where node 4 is the shifted one. The remaining parameters are symmetric, so we consider $p=1\%$ and $T = 100$ ms for all qubits. First, we investigate the optimal measurement pattern for the specified parameters to find that the optimal is always $(\sigma_x, \dots, \sigma_x)$, which is in agreement with the results found for the symmetric scenario. Using this measurement pattern we study the effect of this position shifting on the fidelity of the final Bell pair. 

For the local protocol, the shifting of node 1 highly impacts the final fidelity, as this is the one that starts the chain of measurements, and if it is more separated (close) to the rest of the chain the delay times are overall longer (shorter). However, if the shifted qubit is the other target, the fidelity is not affected since the delay time of qubit 1 is longer. The fidelity increases (decreases) as any of the inner neighbors is shifted closer (farther) from qubit 1. This change in fidelity is the same for any qubit in $\{2, \dots, N-2\}$ as the distance that the request has to be relayed in the network is not altered by the shifting of any of these qubits. However, the shifting of qubit $N-1$ changes the distance of the relaying and thus has a higher impact on the fidelity. 

For the central protocol, the final fidelity is generally higher (lower) the closer (farther) the shifted qubit is from the coordinator, as delay times are shorter (longer) and the impact of the time-dependent dephasing noise is lower (higher). However, if the shifted qubit is a target qubit, then the fidelity does not increase by being closer to the coordinator, as target qubits need to wait for each other. Finally, by shifting the coordinator the fidelity only decreases, because, even though one of the targets then has shorter delay times, the other has longer delay times. 

\begin{figure}[h] 
    \centering
    \subfloat[Local protocol]{\includegraphics[width=0.5\columnwidth]{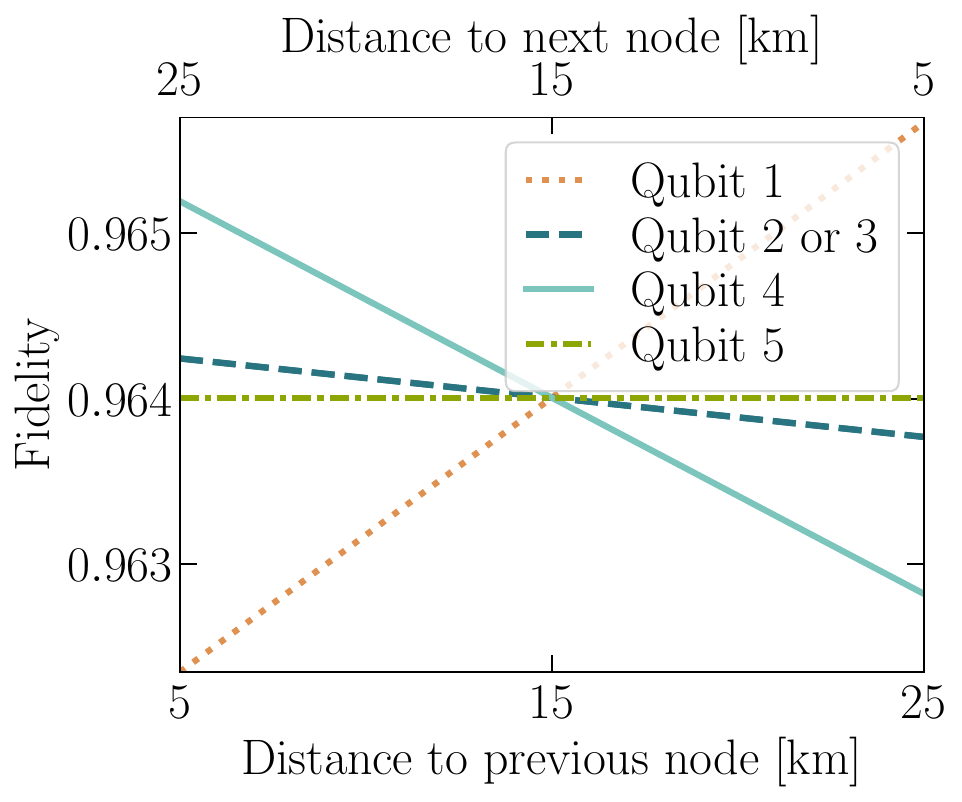} \label{fig:position:asymmetry:local}}
    \subfloat[Central protocol]{\includegraphics[width=0.5\columnwidth]{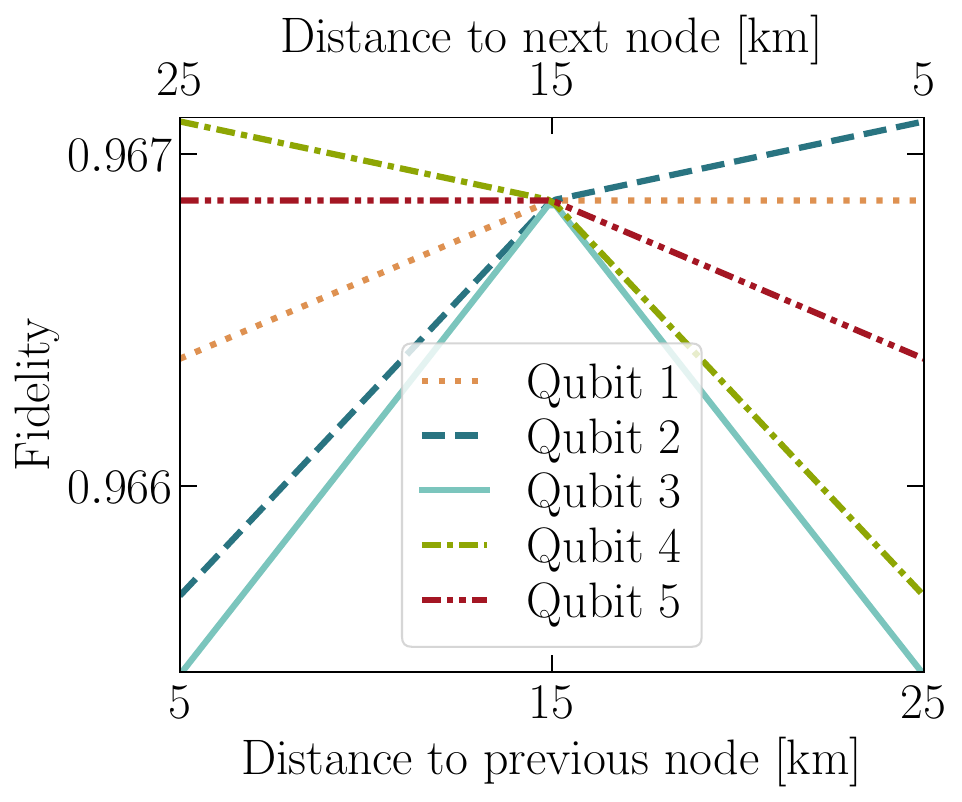} \label{fig:position:asymmetry:central}}
    \caption{Fidelity of a Bell pair obtained from a 5-qubit 1D cluster in terms of the shifted position of one of the nodes with the local or the central protocol. Each of the lines denotes which is the shifted qubit. The bottom $x$ axis denotes the distance to the previous node and the top is the corresponding distance to the next node. Lines that coincide in (a) are plotted in the same style and color.}
    \label{fig:position:asymmetry}
\end{figure}

Fig.~\ref{fig:position:asymmetry} shows the described behavior for both protocols with $N=5$, where we see that the changes in fidelity due to asymmetries in position are relatively small, not even reaching $1\%$. 

\subsection{Custom resource state}
Here we present and analyze an imperfect scenario with asymmetric memories and positions for both local and central protocols using both a basic and a tailored distribution of the resource state. Let us define the number of \textit{hops} as the number of edges between the two target qubits in the resource graph state, i.e., in the entanglement topology. Then, given a set of possible requested bipartite connections in the network, we consider a custom entanglement topology as one that minimizes the number of hops between the nodes to request these bipartite connections. Therefore, this tailoring minimizes the number of qubits to be measured to achieve a desired Bell pair, leading to less noise on the final target state.

\begin{figure}[h]
    \centering
    \includegraphics[width=\columnwidth]{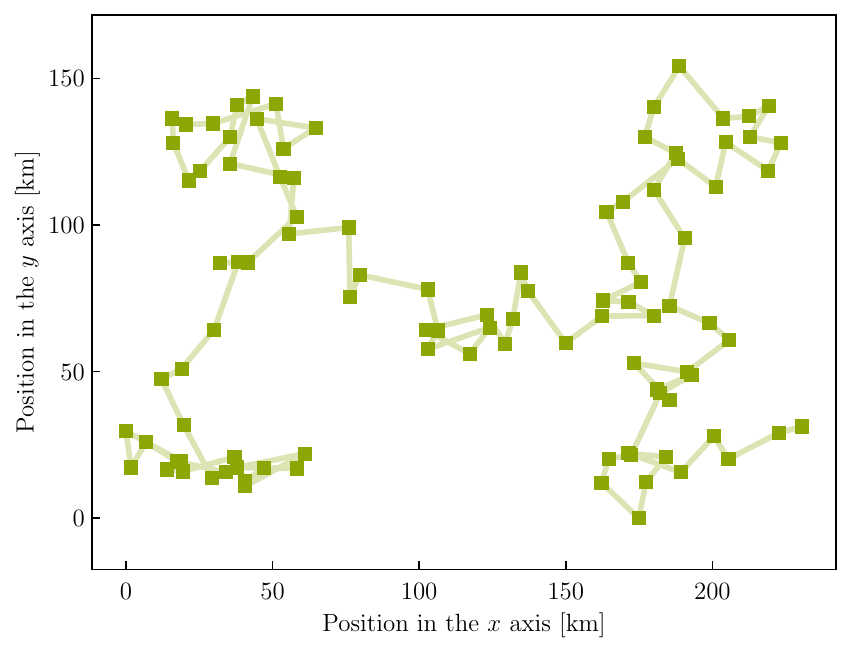}
    \caption{Random positioning of 100 nodes on a flat surface, the line connecting dots denotes neighboring stations.}
    \label{fig:node:position}
\end{figure}

We consider a hundred nodes distributed on a flat surface following a 1D chain with inter-node distance ranging randomly from 5 km to 25 km, as depicted in Fig.~\ref{fig:node:position}. We fix the strength of depolarizing noise to $p=1\%$ and vary randomly the dephasing times such that $T_v\in[10, 100]$ ms for each qubit.

Here, to define the custom entanglement topology, we compute a random distribution of the 1D cluster state, and we assume that all the pairs of nodes that have qubits separated by a low number of hops are the possible nodes to request a bipartite connection. 

To compare the basic and the custom entanglement topologies, we first compute the average fidelity over hops for the custom case. Then, we take the desired pairs, which are a certain number of hops away in the custom structure, and we compute their fidelity in the basic entanglement topology and average them. For both cases, we use the measurement pattern $(\sigma_x, \dots, \sigma_x)$ for the inner neighbors, because as shown in Fig.~\ref{fig:optimal:pattern}, the parameter space we have chosen lies in the depolarizing regime and thus, this measurement pattern has a better behavior. Importantly, we always consider that the target qubits are not at the ends of the cluster, such that all of the computed fidelities include the measurement of the two outer neighbors. 

\begin{figure}[h] 
    \centering
    \subfloat[Local protocol]{\includegraphics[width=\columnwidth]{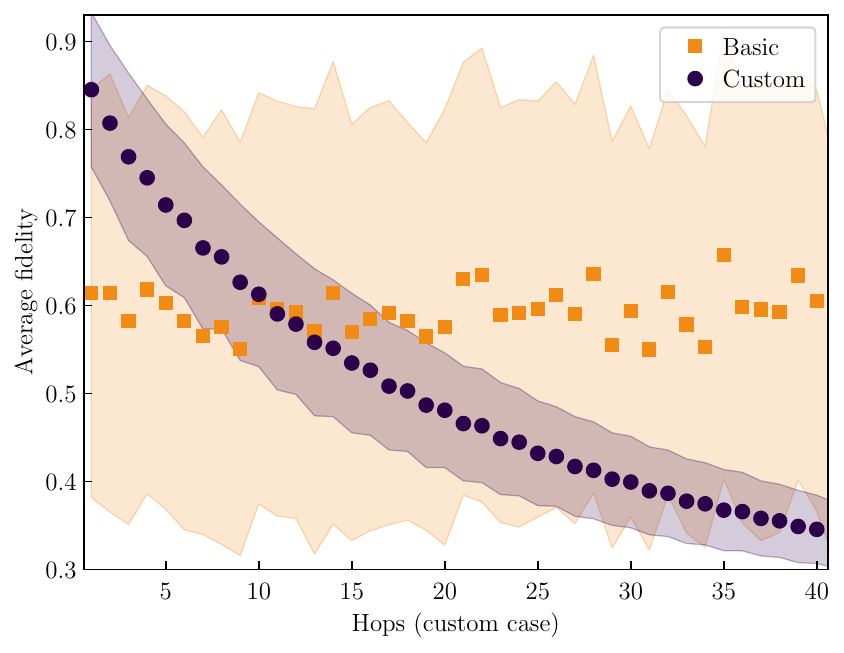}\label{fig:custom:vs:basic:local}}
    \vfill
    \subfloat[Central protocol]{\includegraphics[width=\columnwidth]{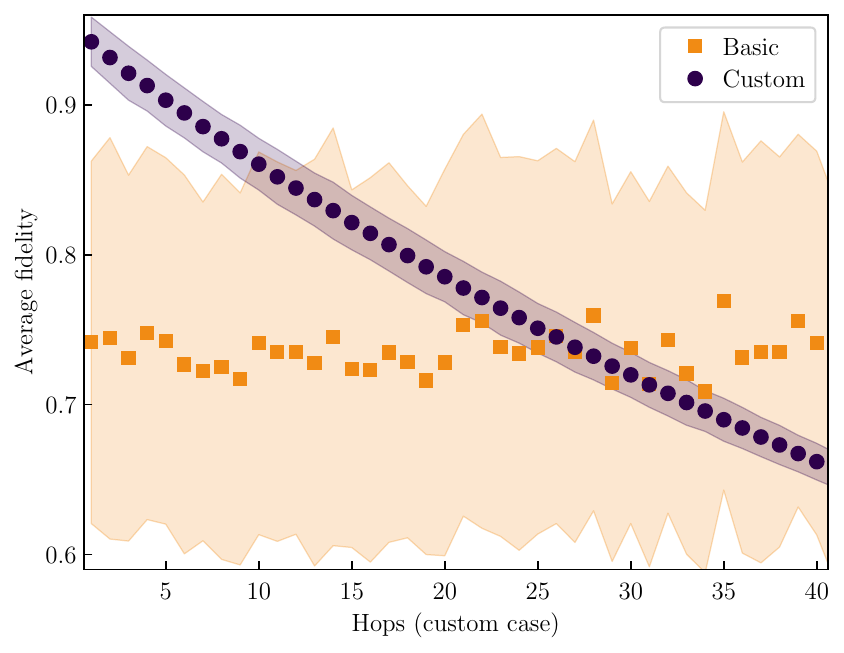}\label{fig:custom:vs:basic:central}}
    \caption{Average fidelity over the hops defined by the custom entanglement topology of a 100-qubit 1D cluster, for both the local and central protocols. The shadowed area shows the corresponding standard deviation.}
    \label{fig:custom:vs:basic}
\end{figure}

In Fig.~\ref{fig:custom:vs:basic}, we plot the results corresponding to the described average fidelities for both protocols. In general, the custom entanglement topology shows a clear advantage regarding the average fidelity for a small number of hops. However, the regime of hops where there is an advantage for the local protocol is smaller than for the central protocol. This is due to the two competing effects present in the local protocol. First, the effect of the initial depolarizing noise is stronger as the number of hops increases. Second, the effect of the time-dependent dephasing noise is minimized by having all the involved parties, both targets and nodes that have to perform measurements, as close as possible to minimize delay times. Therefore, the impact of the initial noise is minimized with the custom resource state, but the impact of the time-dependent noise is minimized using the basic entanglement topology. In contrast, in the central protocol, there are no competing effects, as for both the custom and the basic entanglement topologies the influence of the time-dependent noise is equal. So, minimizing the number of hops achieves higher fidelities. Thus, using a custom entanglement topology with the central protocol is even more advantageous.

Note that the custom entanglement topology in this scenario is chosen randomly and is not related to the geometry of the network. Consequently, the resulting number of hops in the basic entanglement topology varies greatly as we investigate a fixed number of hops in the custom case. Therefore, it is unsurprising that the results for the basic entanglement topology in Fig.~\ref{fig:custom:vs:basic} stay relatively constant with large standard deviations, as the average is taken over many different outcome fidelities resulting from a large variety of hops. This contrasts with the custom case, where the average fidelity decreases as the number of hops increases. 

\begin{figure}[h]
    \centering
    \includegraphics[width=\columnwidth]{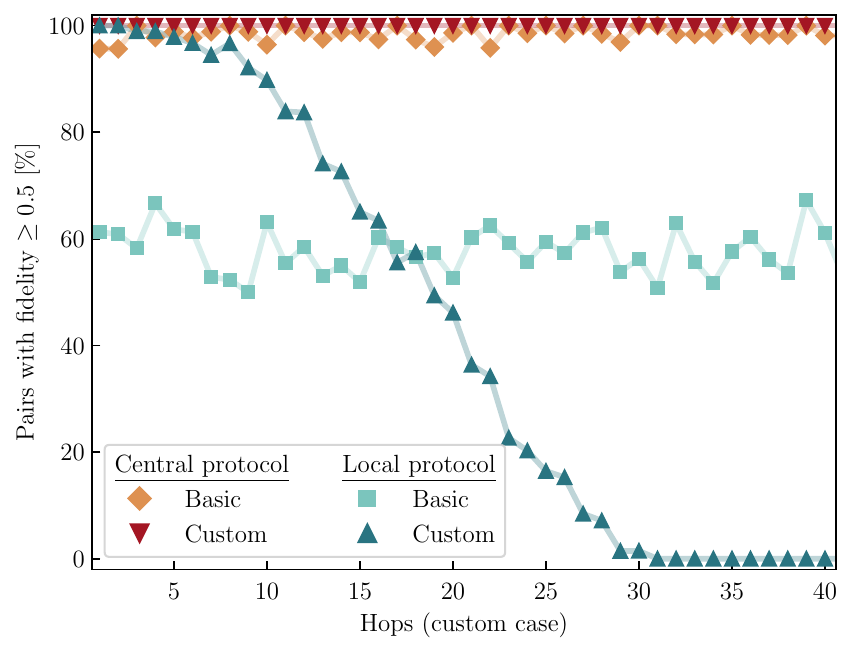}
    \caption{Percentage of final pairs with fidelity greater or equal to 0.5 in terms of the hops defined by the custom entanglement topology of a 100-qubit 1D cluster, for both the local and central protocols.}
    \label{fig:usable:pairs}
\end{figure}

In Fig.~\ref{fig:usable:pairs}, we show the percentage of usable pairs, meaning that their fidelity is greater or equal to 0.5, as a Bell pair with a lower fidelity than 0.5 is no longer entangled. In general, the central protocol has a better performance, which is further enhanced by the custom entanglement topology. This is because the delay times overall in the central protocol are shorter and consequently, the time-dependent noise has less influence on the target state. Nevertheless, note that the performance of the local protocol is greatly improved by the custom entanglement topology for a small number of hops.  

\section{Conclusions and outlook}\label{sec:Conclusion}
We have focused on simulating, optimizing, and tailoring the adaptive phase in imperfect EBQNs, where we consider a cluster state as a resource state and a Bell pair as a target state. In particular, we show that optimizing the measurement pattern used for this manipulation is not only relevant but also the optimal measurement pattern changes depending on the noise parameters in a symmetric scenario. We take this optimization further by considering asymmetries in the dephasing times of the qubits and the positioning of the network nodes, to find that the optimal measurement pattern does not change for the particular parameter space we analyze. Moreover, when considering these asymmetric scenarios we can quantify the impact of each asymmetry in terms of the fidelity of the target state. Lastly, we demonstrate that our method based on the NSF can consider large graph states (up to 100 qubits in our results, but also states of several thousand or tens of thousands of qubits can be simulated) in a highly asymmetric scenario, given our noise model and resource state. Here we show that tailoring both the measurement pattern and the entanglement topology to the type of requests that are expected in the network is of importance due to its advantage in terms of the noise on the final target state. In particular, the use of such tailored resource states has been shown to have a significant effect on (average) reachable fidelities.

We have argued that the 1D cluster as a resource state is a building block for higher-dimensional clusters. However, to fully unleash the possibilities of EBQNs, considering higher-dimensional clusters is crucial, as the large connectivity of these clusters allows for parallel target states (this is limited in the one-dimensional instance). Nonetheless, establishing parallel connections in EBQNs is at its early stages, and only results in the noiseless case have been studied \cite{freund_routing}, which makes it even more appealing to study in an imperfect setting as the one we use here. Nevertheless, in order to consider time-dependent noise with these states, the proposed protocols must be adapted to include network routing \cite{Caleffi2017, hahn_quantum_2019, leung_butterfly, gyongyosi2017entanglement} (which is not needed for the one-dimensional instance), since choosing the shortest path in high-dimensional clusters in an asymmetric setting is not trivial. Moreover, network routing can also be required to expand the local protocol, e.g., considering that each node only knows about the nodes classically connected to it. Additional networking aspects, such as which requests of target states can be fulfilled simultaneously, would need to be considered in this case.

Besides considering high-dimensional clusters, one can also investigate other families of graph states, e.g., tree states, which have been shown to have extremely good behavior under depolarizing noise \cite{morruiz2023influence} at the cost of destroying large parts of the cluster. Moreover, when considering other entanglement topologies one might also want to change the geometry of the network and explore other possibilities of the network protocols. Note that our approach can be used as an inspiration for these new protocols and, moreover, our simulation tools can be freely used to analyze them. 

Nevertheless, looking at more elaborated resource states complicates the customization of the entanglement topology, as the possibilities of such customization become very broad. Therefore, we believe that finding good criteria for customizing entanglement topology based on a set of possible target states for simple resource states, e.g., 1D cluster, is essential to generalize or find good heuristic methods for more complicated resource states. 

Focusing on the dynamic phase where these custom entanglement topologies are built is also relevant, as this generation and distribution process is not an easy task. This analysis could provide a cost function to compare the different entanglement topologies more fairly or it could be also a potential figure of merit to compare the above-mentioned different methods or criteria to customize a certain entanglement topology.

The code we have used for all the results in this work is archived at \cite{code_github}. An implementation of the NSF for graph states is available as a Python package at \cite{nsf_github}.

\section*{Acknowledgments}
M. F. M. R. and W. D. acknowledge that this research was funded in whole or in part by the Austrian Science Fund (FWF) 10.55776/P36010 and 10.55776/P36009. For open access purposes, the author has applied a CC BY public copyright license to any author-accepted manuscript version arising from this submission. Finanziert von der Europäischen Union - NextGenerationEU. J. W. acknowledges support from the Federal Ministry of Education and Research (BMBF) via the project QR.X.
\bibliographystyle{quantum}
\bibliography{refs_Q.bib}
\clearpage
\onecolumn
\renewcommand\appendixname{Appendix}
\appendix

\section{Measurement patterns} \label{a:MeasurementPatterns}
In this section we aim to show explicitly the equivalence between strategies introduced in \cite{mor_noisy} and measurement patterns presented in Sec.~\ref{sec:Setting}. We also present the required tools to translate one to another, and lastly, we analyze the treatment of noisy graph states manipulated via a measurement pattern.

\subsection{Preliminaries}
Here we present the local Pauli measurements on graph states introduced in \cite{hein_multiparty_2004, hein_entanglement_2006}. Consider $v$ to be the qubit of graph state $\left|G\right\rangle$ to be measured in a Pauli basis, then the following unitaries are defined: 
\begin{align}
  U_{z,+}^{(v)} & = \mathds{1}, & U_{z,-}^{(v)} & = \prod_{w\in \mathcal{N}_v} \sigma_z^{(w)}, \label{eq:correction:z}\\
  U_{y,+}^{(v)} & = \prod_{w\in \mathcal{N}_{v}}\sqrt{-i \sigma_z^{(w)}}, & U_{y,-}^{(v)} & = \prod_{w\in \mathcal{N}_v}\sqrt{i \sigma_z^{(w)}}, \label{eq:correction:y}
\end{align}
and, depending furthermore on a vertex $w_0\in \mathcal{N}_v$,
\begin{equation} \label{eq:correction:x}
     U_{x,+}^{(v)}= \sqrt{i \sigma_y^{(w_0)}}\prod_{w\in \mathcal{N}_v-\mathcal{N}_{w_0}-\{w_0\}}\sigma_z^{(w)},\quad\qquad 
     U_{x,-}^{(v)} =\sqrt{-i \sigma_y^{(w_0)}}\prod_{w\in \mathcal{N}_{w_0}-\mathcal{N}_v-\{v\}}\sigma_z^{(w)}.
\end{equation}
After the measurement of qubit $v$ in a Pauli basis, $\alpha =x, y, z$, the state, depending on the outcome $\pm 1$, is given by
\begin{equation}\label{eq:pauli:measurement}
P_{\alpha,\pm}^{(v)}\left|G\right\rangle = \left|\alpha,\pm\right\rangle^{(v)}\otimes U_{\alpha,\pm}^{(v)}\left|G'\right\rangle
\end{equation}
where $P_{\alpha,\pm}^{(v)}$ is the projector of qubit $v$ in the $\alpha$ basis, such that  $\left|\alpha,\pm\right\rangle^{(v)}$ is the state of qubit $v$ after the measurement, which is the eigenvector with eigenvalue $\pm 1$ for the corresponding basis. Note that for the $\sigma_x$ measurement, the unitary depends on the choice of $w_0$. However, the resulting graph states arising from different choices of $w_0$ and $w'_0$ will be equivalent up to the local unitary $U_{w'_0}^{\tau}\left(U_{w_0}^{\tau}\right)^{\dagger}$, where 
\begin{equation}
    U_{w}^{\tau} = \sqrt{-i\sigma_x^{(w)}}\prod_{u\in \mathcal{N}_w}\sqrt{i \sigma_z^{(u)}}
\end{equation}
denotes the local complementation on $w$. 

The resulting graph $G'$ after the measurements is given by the rules described in Sec.~\ref{sec:Concepts}, and after applying the corresponding correction operator Eq.~\eqref{eq:correction:z}, Eq.~\eqref{eq:correction:y} or Eq.~\eqref{eq:correction:x}, we write the resulting graph state as
\begin{equation}
\left|G'\right\rangle = U_{\alpha,\pm}^{\dagger \, (v)} P_{\alpha,\pm}^{(v)}\left|G\right\rangle,
\end{equation}
where $\alpha = x, y, z$ denotes the basis of the measurement, and for simplicity, we consider always the tracing out of the measured qubit, also, we refer to $U_{\alpha,\pm}^{\dagger \, (v)}$ as the correction operator of the measurement of qubit $v$ in $\alpha$. 

\subsection{Equivalence between strategies and measurement patterns}
\label{a:equivalence}
Consider a graph state $\left|G\right\rangle$ that is manipulated by a set of $n$ measurements to achieve the graph state $\left|G'\right\rangle$. Consider that the strategy is composed by measuring qubits $v_1, v_2, \dots, v_n$ in this order in the following bases $\alpha_1, \alpha_2, \dots, \alpha_n$. After each measurement, the corresponding correction operator must be applied to perform the next measurement in the graph state basis, such that the final graph state is 
\begin{equation}
    \left|G'\right\rangle = U_{\alpha_n,\pm}^{\dagger \, (v_n)}P_{\alpha_n,\pm}^{(v_n)} \cdots U_{\alpha_1,\pm}^{\dagger \,(v_1)} P_{\alpha_1,\pm}^{(v_1)} \left|G\right\rangle.
\end{equation}
Given the fact that the correction operators are Clifford and projectors are on a Pauli basis, we can commute to the left the correction operators, such that
\begin{equation}
    \left|G'\right\rangle = U_{\alpha_n,\pm}^{\dagger \, (v_n)} \cdots U_{\alpha_1,\pm}^{\dagger \,(v_1)} P_{\alpha'_n,\pm'}^{(v_n)} \cdots P_{\alpha_1,\pm}^{(v_1)} \left|G\right\rangle.
\end{equation}
Note that this commutation is such that the bases and outcomes of the projectors for $i\in\{2,\dots,n\}$ might change to another Pauli basis, such that  
\begin{equation} \label{eq:pattern}
    P_{\alpha'_i,\pm'}^{(v_i)} = \left(U_{\alpha_1,\pm}^{(v_1)} \cdots U_{\alpha_{i - 1}, \pm}^{(v_{i - 1})}\right) P_{\alpha_i ,\pm}^{(v_i)} \left(U_{\alpha_{i-1},\pm}^{\dagger \, (v_{i-1})} \cdots U_{\alpha_1,\pm}^{\dagger \,(v_1)}\right).
\end{equation}
Importantly, no correction operator needs to commute with the projector for the first measurement, the one on qubit $v_1$, and thus, $\alpha'_1 = \alpha_1$. We denote the measurement pattern as $\mathcal{P}=P_{\alpha'_n,\pm'}^{(v_n)} \cdots P_{\alpha_1,\pm}^{(v_1)}$ and the overall correction operator as $\mathcal{U} = U_{\alpha_n,\pm}^{\dagger \, (v_n)} \cdots U_{\alpha_1,\pm}^{\dagger \,(v_1)}$. 

Therefore, the strategy of measuring qubits $v_1, v_2, \dots, v_n$ in this order in these bases $\alpha_1, \alpha_2, \dots, \alpha_n$, correspondingly, is equivalent to applying $\mathcal{P}$, where all elements commute with each other, followed by $\mathcal{U}$, which can be applied in a post-processing step.

Notice, that this equivalence also holds for a noisy graph state, as the commutation of the correction operators with the measurement projectors does not change if they are applied to a graph state subject to noise. 

\subsection{Translation between strategies and measurement patterns}
Appendix~\ref{a:equivalence} already gives the general way to calculate the measurement bases for the measurement pattern corresponding to a given NSF strategy. However, since there are only a few possible operators and measurement bases, one can construct a straightforward algorithm to determine the bases of the measurement pattern.

The key commutation relations are
\begin{equation}
    \begin{aligned}
        \sigma_x \sqrt{-i\sigma_z} &= - \sqrt{-i\sigma_z} \sigma_y, \\
        \sigma_y \sqrt{-i\sigma_z} &= + \sqrt{-i\sigma_z} \sigma_x, \\
        \sigma_x \sqrt{+i\sigma_y} &= - \sqrt{+i\sigma_y} \sigma_z,\\
        \sigma_z \sqrt{+i\sigma_y} &= + \sqrt{+i\sigma_y} \sigma_x.
    \end{aligned}
\end{equation}
Together with equations \eqref{eq:correction:y} and \eqref{eq:correction:x} it follows that a $\sigma_y$ measurement on a qubit $v$ leads to an exchange of future $\sigma_x$ and $\sigma_y$ measurements on the neighborhood of $v$. Furthermore, a $\sigma_x$ measurement on a qubit $v$ leads to an exchange of later $\sigma_x$ and $\sigma_z$ measurements on the special neighbor $w_0$.

For the purposes of this discussion, we do not keep track of the sign changes of the measurements, since they are unimportant for how the graph state transforms. However, to actually perform the protocol one needs to take this into account as well by following Eq.~\eqref{eq:pattern}.

Consider an initial graph $G$ with $N$ vertices and an ordered sequence of $n$ graph transformation instructions $\{S_j\}_{j=1}^n$. Each element $S_j$ consists of a measurement basis $\alpha_j$, where $v_j$ is the label of the qubit to be measured (for measurements in $\sigma_x$ basis the label of the special neighbor is denoted by $w_{0,j}$). 

We calculate the sequence of graphs that result from successively applying the instructions and using the graph update rules (with corrections) $\{G_j\}_{j=1}^n$, where $G_l$ denotes the graph after having applied $\{S_j\}_{j=1}^{l-1}$ successively to the initial graph $G$.  Note that $G_1 = G$, as no transformation has been applied yet. Then, $\mathcal{N}(G_j)_v$ denotes the neighborhood of vertex $v$ according to the graph $G_j$.

Initialize a vector $\bm{\mu}$ of length $N$, where the $i^{\text{th}}$ entry denotes if the $i^{\text{th}}$ qubit would be measured according to the sequence $\{S_j\}$ and if so, in which basis, such that 
\begin{equation}
    \mu_i = \begin{cases}
            \sigma_{\alpha_k} &\text{ if } \exists\, k \text{ s.t. } i = v_k, \\
            \mathds{1} &\text{ otherwise,}
           \end{cases}
\end{equation}
where $\mathds{1}$ denotes that the $i^{\text{th}}$ qubit is never measured.

Now, we iterate over the sequence $\{S_j$\} \textit{in reverse order} and update $\bm{\mu}$ as following these rules, where $j$ denotes which is the $S_j$ being applied:
\begin{equation}
    \begin{aligned}
    \text{If } \alpha_j = & z: \\
                          & \mu^\prime_i = \mu_i.   \\
    \text{If } \alpha_j = & y: \\
                          & \mu^\prime_i  = 
                            \begin{cases}
                                \sigma_x & \text{ if } \mu_i = \sigma_y \text{ and } i \in \mathcal{N}(G_j)_{v_j}, \\
                                \sigma_y & \text{ if } \mu_i = \sigma_x \text{ and } i \in \mathcal{N}(G_j)_{v_j}, \\
                                \mu_i    & \text{ otherwise.}
                            \end{cases} \\
    \text{If } \alpha_j = & x: \\
                          & \mu^\prime_i = 
                            \begin{cases}
                                \sigma_x & \text{ if } \mu_i = \sigma_z \text{ and } i = w_{0,j}, \\
                                \sigma_z & \text{ if } \mu_i = \sigma_x \text{ and } i = w_{0,j}, \\
                                \mu_i    & \text{ otherwise.}
                            \end{cases}
    \end{aligned}
\end{equation}
Then, let $\bm{\mu}^\prime \rightarrow \bm{\mu}$ and continue the iteration. The final $\bm{\mu}$ is then the measurement pattern that is equivalent to the sequence of instructions $\{S_j\}$.

Each sequence of graph transformations corresponds to a single measurement pattern, but a measurement pattern can correspond to multiple possible graph transformation sequences. One can obtain them, by inverting the procedure above and varying all possible orders of qubits in $\{v_j\}$ as well as all possible choices of the special neighbor $w_0$, if the measurement of the pattern gets mapped to an $x$ graph transformation.

In Sec.~\ref{sec:Results} we use vector $\boldsymbol{m}$ to characterize the measurement pattern of the manipulation of the inner neighbors of a 1D cluster to achieve a Bell pair. This is an instance of the general vector $\bm{\mu}$ presented here. Note that $\bm{\mu}$ and the procedure to calculate it is valid for any graph structure and manipulation.

\subsection{Manipulation of noisy graph states with measurement patterns} \label{a:noise}
As mentioned in Appendix~\ref{a:equivalence}, the equivalence between measurement patterns and strategies also holds for a noisy graph state. In this work, where we study EBQN with spatially separated nodes and memory dephasing, the use of measurement patterns is crucial to optimize delay times in the network protocols. Moreover, to simulate and computationally analyze these protocols and the time-dependent noise, the use of the NSF comes as an advantage. Therefore, let us explain how the use of the NSF, based on strategies, with measurement patterns.

First, a strategy is formed by a set of manipulation operators that map a graph state to another one, e.g., a local Pauli measurement followed by the corresponding correction operator. When applying a strategy to a noisy graph state, the NSF gives an update rule on how the noise changes after each manipulation operator, so by applying the sequence of manipulation operations to (i) the graph and (ii) the noise operators, step by step, one can retrieve the final graph and the corresponding noise operators. Note that the update rules have been already computed in \cite{mor_noisy} and coded in \cite{nsf_github} for a set of manipulation operators that are frequently used. 

Now, if one would like to analyze the noise in terms of measurement patterns using the NSF, first one needs to define the manipulation operators that map a graph state to another, this would correspond to applying the measurement pattern and the corresponding overall correction operator afterward, so $\mathcal{U}\mathcal{P}$ (one could view this operator as an overall strategy with a single manipulation operator). Then, one has to compute the update rule of this operator, which corresponds to the collection of all the step-by-step update rules used for an equivalent strategy. Finally, the update rule is applied to the graph and the noise operators, retrieving the final graph and the corresponding noise operators, which would be the same as for an equivalent strategy. 

The difference between using the NSF with strategies or measurement patterns is that for each measurement pattern, a unique update rule has to be computed and for the strategy - since one goes manipulation operator by manipulation operator - the update rules are already pre-calculated and can be reused for different strategies. Nevertheless, both approaches lead to the same final graph and noise operators. 

Therefore, in terms of the simulation of noisy graph states, the strategies approach is more efficient. Nevertheless, in our work, we use measurement patterns to describe the timing effects of the network. To simulate it we take the considered measurement pattern, translate it to one of the equivalent strategies, and use that.

As mentioned above, a measurement pattern can correspond to multiple strategies. This also holds in terms of noise, meaning that the collections of all the update rules of each of these strategies lead to the same overall update rule. Then this collection would be the one corresponding to the manipulation operator $\mathcal{U}\mathcal{P}$, where $\mathcal{P}$ is the measurement pattern all strategies are equivalent to. 

The fact that several strategies might translate to the same measurement pattern is key for the search for the optimal measurement pattern to manipulate an $N$-qubit 1D cluster into a Bell pair between the ends presented in Sec.~\ref{ssec:Optimal}. All possible measurement patterns are $2^{N-2}$, as one has to count all the possible combinations of $\sigma_y$ and $\sigma_x$ measurements. Note that, in this case, there is no measurement pattern containing $\sigma_z$, since all measurements in a measurement pattern commute, such that a $\sigma_z$ measurement could be performed first, leading to a cut in the 1D cluster that would prevent the manipulation to achieve a Bell pair between the ends. In order to perform the exhaustive search, we take an equivalent strategy for each possible measurement pattern and we simulate the corresponding scenario. Importantly, here the use of measurement patterns is again an advantage, because if we searched directly over possible strategies we would have to do so over $3^{N-2}\cdot (N-2)!$ strategies (neglecting the choice of the special neighbor for the $\sigma_x$ measurement) and some would not retrieve a Bell pair, but two not entangled qubits instead.

\section{Delay times in entanglement-based quantum network protocols}\label{a:delay:times}
Here we compute the delay times for all the involved qubits in both the local and central protocols.

\subsection{Local protocol}\label{a:delay:times:local}
For a certain request, we define the target nodes to establish a Bell pair between, $a$ and $b$, and the qubits that have to be measured, $\{q_i\}$. Let us define $\{n_i\}$ as the nodes that have a qubit that belongs to $\{q_i\}$. Next, let us define $d_{\xi, n_i}$ as the path distance between a node in $\{n_i\}$ and a target node $\xi = a, b$ as the sum of the lengths of the classical communication channels between them. The delay time for a qubit in $\{n_i\}$ is 
\begin{equation}
    \frac{d_{a, n_i}}{\nu} + \tau_{n_i},
\end{equation}
where $\tau_{n_i}$ is the processing time of qubit in $n_i$ and $\nu$ is the communication speed, such that $d_{a, n_i}/\nu$ is the time for the request started by $a$ to be relayed by all the necessary nodes. As mentioned in Sec.~\ref{sec:Setting}, in the manipulation there are the inner and outer neighbors. Since each of the two outer neighbors is connected to either $a$ or $b$, we label them as $n_a$ and $n_b$, and we consider $l$ inner neighbors so that we label them as $\{n_1, n_2, \dots, n_l\}$. Here, the delay time of the targets includes (i) the time to relay the request to all involved nodes, (ii) the time to perform the measurements and send the outcomes to both $a$ and $b$, and (iii) the time for the other target to finish (i) and (ii). The time to complete (i) and (ii) is
\begin{equation}
   g_{\xi} = \max_{i\in\{\xi, 1, 2, \dots, l\}}\left(\frac{d_{a, n_i} + d_{\xi, n_i}}{\nu} + \tau_{n_i}\right) ,
\end{equation}
where $\xi = a, b$. The target delay time, which includes (iii), is the maximum between $g_{a}$ and $g_{b}$.

\subsection{Central protocol}\label{a:delay:times:central}
We define the coordinator $C$, and for a certain request, the target nodes to establish a Bell pair between are denoted by $a$ and $b$, and the nodes that have to measure to achieve the target are $\{n_i\}$. 
For any qubit in $\{n_i\}$ the delay time is
\begin{equation}
    \frac{d_{n_i}}{\nu} + \tau_{n_i},
\end{equation}
where $d_{n_i}$ is the length of the classical communication channel between $C$ and $n_i$, $\nu$ is the communication speed, and $\tau_{n_i}$ is the processing time of $n_i$. As in the previous protocol, there are two outer neighbors, $n_{a}$ and $n_{b}$, where each is connected to either $a$ or $b$, respectively, furthermore, there are $l$ inner neighbors, which we denote as $\{n_1, n_2, \dots, n_l\}$. So, the delay time of the targets includes (i) the time to send all commands and perform them, (ii) the time to receive the correction operator, and (iii) the time for the other target to finish (i) and (ii). The time to complete (i) and (ii) is
\begin{equation}
    f_{\xi}= \frac{d_{\xi}}{\nu} + \max_{i\in\{\xi, 1, 2, \dots, l\}}\left(\frac{2d_{n_i}}{\nu} + \tau_{n_i}\right) ,
\end{equation}
where $\xi = a, b$ and $d_{\xi}$ the length of the classical communication channel between $C$ and $\xi$. The target delay time, which includes (iii), is the maximum between $f_a$ and $f_b$.

\section{Achievable fidelities in a symmetric scenario}\label{a:symmetric}
\begin{figure}[h] 
    \centering
    \subfloat[Local protocol]{\includegraphics[width=0.5\columnwidth]{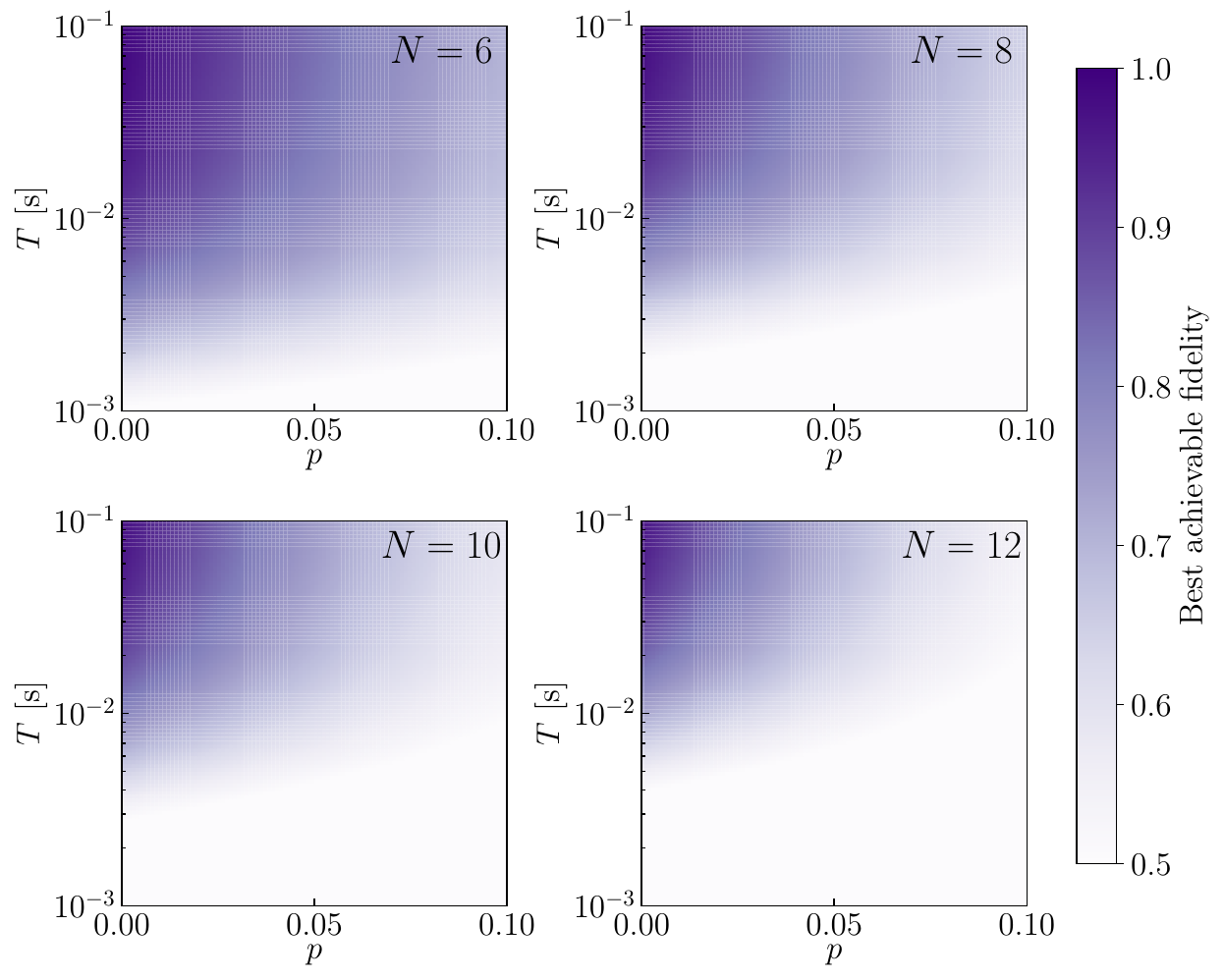}}
    \subfloat[Central protocol]{\includegraphics[width=0.5\columnwidth]{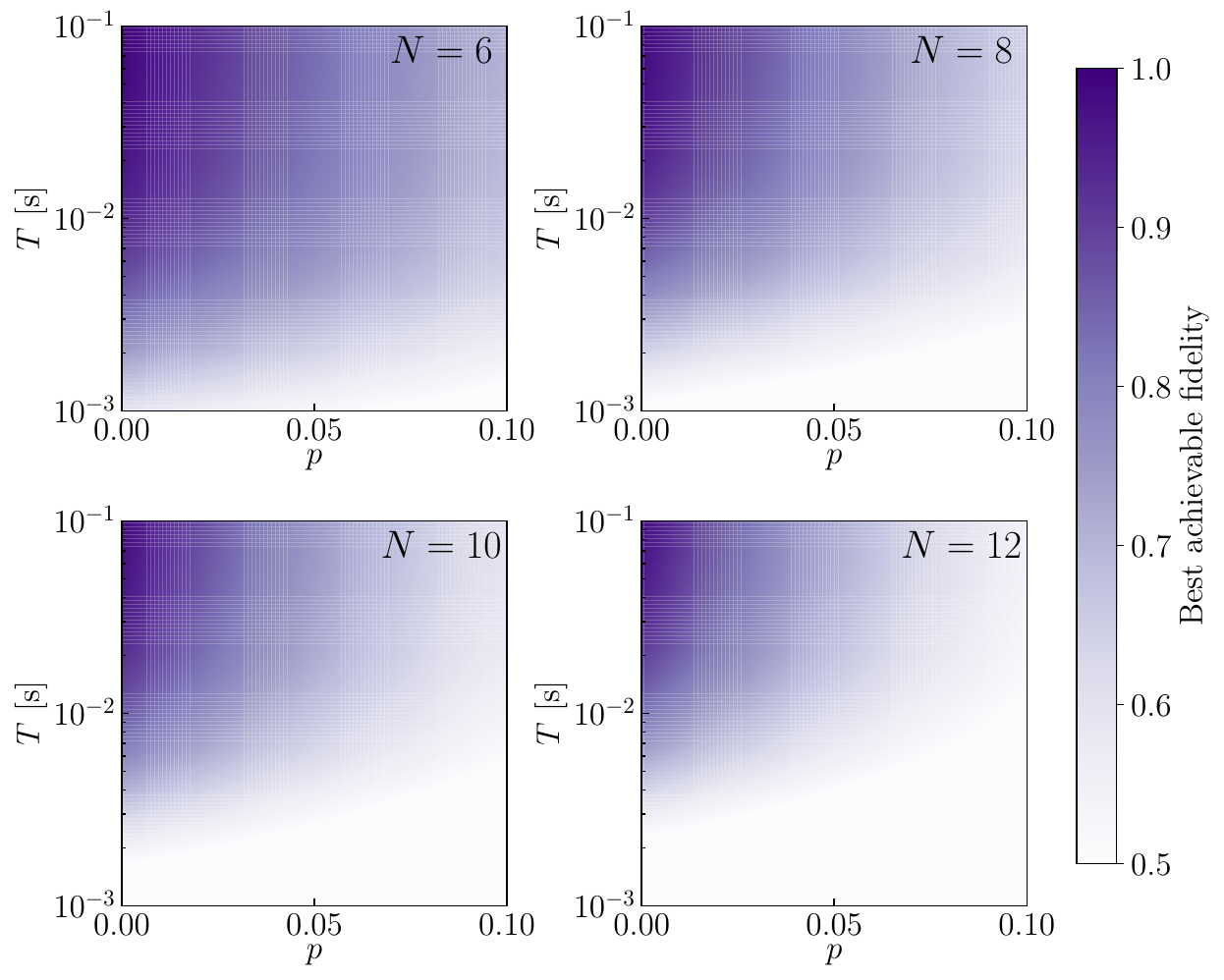}}
    \caption{Best achievable fidelity for the local and the central protocol, with even $N$, in terms of the strength of the depolarizing noise $p$ and the dephasing time $T$. The white area denotes that the fidelity is below 0.5, so the target Bell pair is no longer entangled.}
    \label{fig:best:fidelity}
\end{figure}

Considering the scenario described in Sec.~\ref{ssec:Optimal}, in Fig.~\ref{fig:best:fidelity}, we plot the fidelity of a Bell pair achieved via the manipulation of $N$-qubit 1D cluster, where the inner neighbors are measured using the optimal measurement. Note that depending on the strength of the depolarizing noise and the dephasing time, the chosen measurement pattern changes according to the results presented in Sec.~\ref{ssec:Optimal}.

\end{document}